\documentclass[twocolumn,showpacs,pre,floatfix,amsmath,amssymb,notitlepage]{revtex4-1}
\usepackage{epsfig}
\usepackage{subfigure}
\usepackage{amsmath}
\usepackage{color}
\usepackage{amssymb}
\usepackage{setspace}
\usepackage{graphicx}% Include figure files
\usepackage{dcolumn}% Align table columns on decimal point
\usepackage{bm}% bold math
\usepackage{times}
\usepackage{enumerate}
\usepackage{footnote}
\usepackage{upgreek}
\input epsf

\graphicspath{{figures/}}

\begin{document}

	\author{Linpeng Gu$^{1}$}
	\author{Liang Fang$^{1}$} 
	\author{Hanlin Fang$^{2}$}
	\author{Juntao Li$^{2}$} 
	\author{Jianbang Zheng$^{1}$}
	\author{Jianlin Zhao$^{1}$} 
	\author{Qiang Zhao$^{3*}$, Xuetao Gan$^{1}$} 
	\email{zhaoqiang@qxslab.cn; xuetaogan@nwpu.edu.cn} 
	\affiliation{$^{1}$MOE Key Laboratory of Material Physics and Chemistry under Extraordinary Conditions, and Shaanxi Key Laboratory of Optical Information Technology, School of Science, Northwestern Polytechnical University, Xi'an 710072, China}
	\affiliation{$^{2}$State Key Laboratory of Optoelectronic Materials and Technologies, School of Physics, Sun Yat-Sen University, Guangzhou, 510275, China}
	\affiliation{$^{3}$Qian Xuesen laboratory of Space Technology, China Academy of Space Technology, Beijing 100094, China}
	
	\date{\today}
	
	\title{Fano resonance lineshapes in a waveguide-microring structure enabled by an air-hole}

\maketitle

{\bf
We propose and demonstrate, by simply inserting an air-hole in the waveguide side-coupling with a microring resonator (MRR), the transmission spectrum presents Fano lineshapes at all of the resonant modes. Measured from the fabricated devices, Fano lineshapes with slope rates over 400 dB/nm and extinction ratios over 20 dB are obtained. We ascribe it to the air-hole-induced phase-shift between the discrete resonant modes of the MRR and the continuum propagating mode of the bus-waveguide, which modifies their interference lineshapes from symmetric Lorentzian to asymmetric Fano. From devices with varied locations and diameters of the air-hole, different Fano asymmetric parameters are extracted, verifying the air-hole-induced phase-shifts. This air-hole-assisted waveguide-MRR structure for achieving Fano resonance lineshapes has advantages of simple design, compact footprint, large tolerance of fabrication errors, as well as broadband operation range. It has great potentials to expand and improve performances of on-chip MRR-based devices, including sensors, switchings and filters. }\\

\section{Introduction}\label{sec:01}
Microring resonators (MRRs) play crucial roles in on-chip interconnect, signal processing, and nonlinear optics~\cite{del2007optical, guo2017parametric, xu2005micrometre, kippenberg2011microresonator}. Fano resonance lineshapes of MRRs have recently attracted intense interest for improving these chip-integration functions~\cite{wang2016fano, zhao2016tunable, qiu2012asymmetric, hu2013tunable}. As opposed to usual symmetric Lorentzian resonance lineshapes, they have asymmetric and sharp slopes around the resonant wavelengths. The wavelength range for tuning the transmission from zero to one is much narrower in Fano lineshapes, which therefore strengthens the figure of merits of power consumption, sensing sensitivity, extinction ratio, etc.~\cite{mario2006asymmetric, tu2017high, chao2003biochemical, wang2016fano}. Fano resonance is recognized as a general phenomenon in physical waves and originates from the interference between continuum state and discrete localized state~\cite{miroshnichenko2010fano}. By considering resonant modes in MRRs as the discrete state, Fano resonances were realized by coupling MRRs with other photonic structures, including Mach-Zehnder interferometers ~\cite{absil2000compact, zhou2007fano, darmawan2005phase} and Fabry-Perot cavities~\cite{zhang2016optically, liang2006transmission}, which provide a quasi-continuum mode. Unfortunately, these integrated structures sacrifice the compact footprint of MRRs, and it is challenging to achieve precise structure designs and device fabrications for overlapping MMR's discrete modes with the quasi-continuum modes. Also, because of the limited bandwidth of the quasi-continuum modes, Fano resonances only form at certain resonant wavelengths of the MRR, which is incoordinate with the broadband operation of the MRR-based devices. 

Here, we report a compact design to realize Fano lineshapes at all of the resonant modes of a MRR, which also has large design and fabrication tolerances. In the widely reported MRR-based devices, the MRR is normally side-coupled with a bus-waveguide to access its resonant modes, as shown in Fig.~\ref{fig:fig1}(a). Our proposed design is inserting an air-hole in the bus-waveguide around the waveguide-MRR coupling region, as shown in Fig.~\ref{fig:fig1}(b). This inserted air-hole functions as a phase-shifter between the discrete resonant mode in MRR and the continuum propagating mode in the bus-waveguide, which is experimentally verified by the transmission spectra of the fabricated devices with different air-hole locations and diameters. The obtained slope rate (SR) and extinction ratio (ER) of the Fano resonance lineshapes are exceeding 20 dB and 400 dB/nm, respectively.

\section{Model And Theory}\label{sec:02}
\begin{figure}[http]
	\centering
	\fbox{\includegraphics[width=\linewidth]{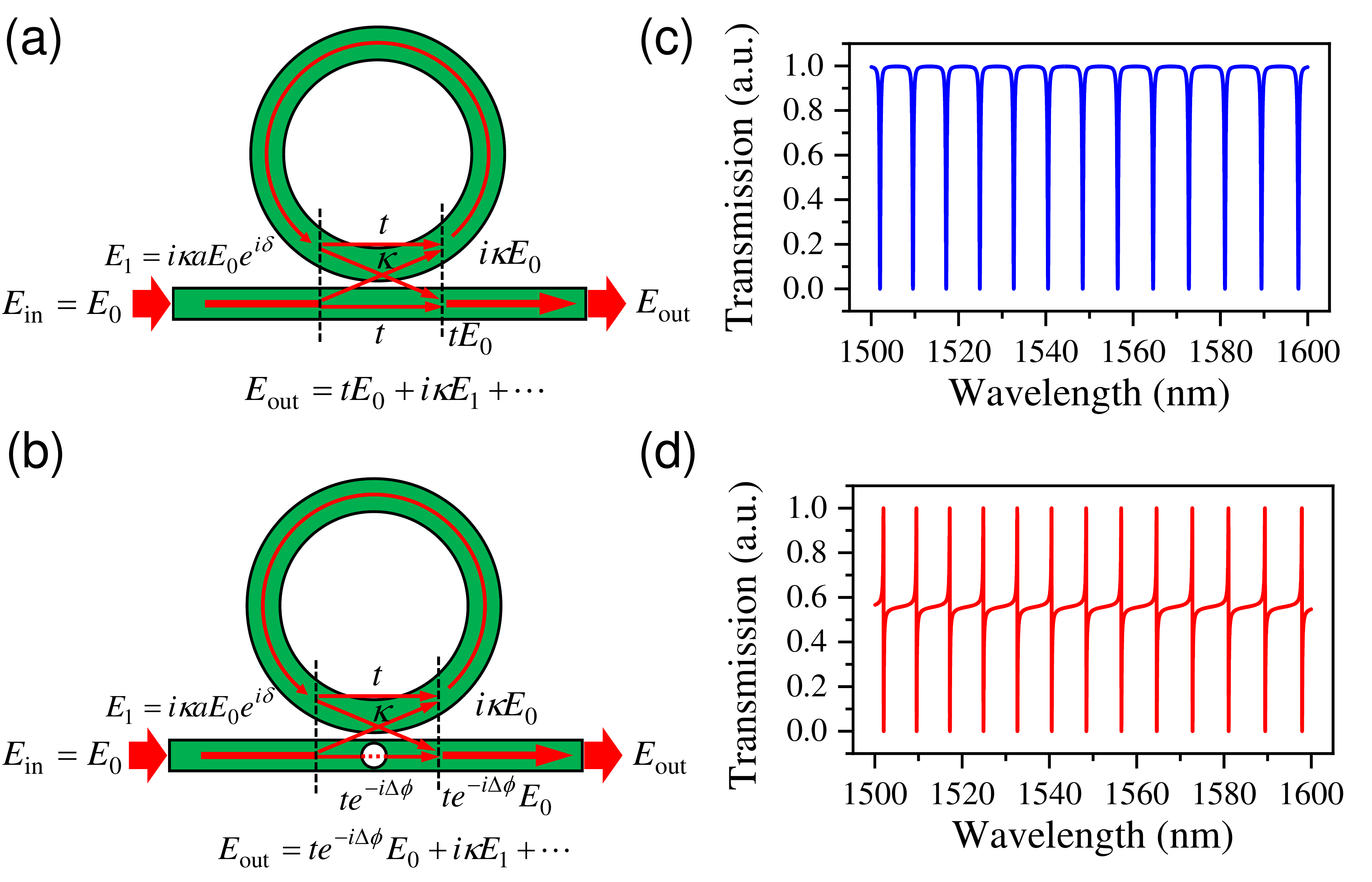}}
	\caption{(a) Propagation of optical filed in a conventional waveguide-MRR structure. (b) Propagation of optical filed in the proposed waveguide-MRR structure with an air-hole inserted in the bus-waveguide. (c) Transmission spectrum calculated by Eq.~\ref{eq:2} with parameter of $t=a=0.95$. (d) Transmission spectrum calculated by Eq.~\ref{eq:4} with parameter of $t=a=0.95$ and $\Delta\phi=\pi/3$.}
	\label{fig:fig1}
\end{figure}

To explain the design principle, we first analyse the propagation of optical field in the conventional waveguide-MRR coupling structure shown in Fig.~\ref{fig:fig1}(a). When an incident optical field $E_{in}=E_0$ propagates in the bus-waveguide, it will split into two parts at the waveguide-MRR coupling region due to the evanescent-field coupling. One of them propagates continuously in the bus-waveguide with an amplitude of $tE_0$ after the coupling region, where $t$ is the transmission coefficient of the coupling region. The other part couples into the MRR with an amplitude of i${\kappa}E_0$, and i${\kappa}$ is the coupling coefficient from one waveguide to the other. This part will propagate along the MRR and circulate back to the coupling region with an amplitude of $E_1=(i\kappa)ae^{i\delta}E_0$, where $\delta={2\pi{nL_R}/\lambda}$ is the round trip phase delay and $a=\exp(-\alpha{L_R})$ is the round trip amplitude transmission coefficient. Here, $n$ is the effective refractive index of the propagating mode, $\lambda$ is the operating wavelength, $L_R$ is the perimeter of the MRR and $\alpha$ is the linear loss coefficient. Then, when the optical field $E_1$ passes through the waveguide-MRR coupling region, with the same coupling mechanism, the part with an amplitude of $tE_1$ continuously propagates in the MRR, and the other part with an amplitude of i${\kappa}E_1$ couples into the bus-waveguide, which will be added with the initial propagating field of $tE_0$ at the output port of the bus-waveguide. For the part of $tE_1$, it will circulate in the MRR and couple partly into the bus-waveguide again at the coupling region. Under this coupling mechanism, the optical fields in the MRR would be added repetitively into the bus-waveguide after each circulation. The total output field of the bus-waveguide could be calculated as 

\begin{equation}
\begin{split}
E_{out}&=\underbrace{t{E_0}}_{\text{continuum state}}+\underbrace{i\kappa{E_1}+i\kappa{E_2}+\cdots}_{\text{discrete state}}\\
&=t{E_0}+i\kappa{(i\kappa)ae^{i\delta}E_0}+i\kappa{(i\kappa)ta^2e^{i2\delta}E_0}+\cdots\\
&=t{E_0}+(i\kappa)^2{\frac{ae^{i\delta}-t^na^{n+1}e^{i(n+1)\delta}}{1-tae^{i\delta}}E_0}~~(t<1,n\rightarrow\infty)\\
&=({t-{\frac{\kappa^2ae^{i\delta}}{1-tae^{i\delta}}})E_0}
\label{eq:1}
\end{split}
\end{equation}

By assuming there is no loss in the waveguide-MRR coupling region, i.e., $t^2+\kappa^2=1$, the final power transmission spectrum of the coupled system is given by 
 \begin{equation}
T(\lambda)=\left|\frac{E_{out}}{E_{in}}\right|^2=\left|\frac{t-ae^{i{2\pi{nL_R}/\lambda}}}{1-tae^{i{2\pi{nL_R}/\lambda}}}\right|^2
\label{eq:2}
\end{equation}

This transmission spectrum has been widely studied and we plot it in Fig.~\ref{fig:fig1}(c) with parameters of $t=a=0.95$ and $nL_R=300.4$ $\upmu$m. It presents periodic Lorentzian dips at the wavelengths that makes phase delay $\delta={2\pi{nL_R}/\lambda}$ equal to the integer multiple of 2$\pi$. At those wavelengths, the optical fields circulating in the MRR satisfy constructive interference and form resonant modes of the MRR. Correspondingly, in Eq.~\ref{eq:1}, the additions after the first term could be considered as a discrete localized state. As discussed above, the first term of Eq.~\ref{eq:1} represents the propagating mode of the bus-waveguide after the coupling region, which is a continuum state of the coupling system. Hence, around the resonant wavelengths, the output field ($E_{out}$) of the waveguide-MRR is an interference of a discrete state and a continuum state. The symmetric Lorentzian transmission dips (see Fig.~\ref{fig:fig1}(c)) of the conventional waveguide-MRR structure shown in Fig.~\ref{fig:fig1}(a) just arise from this interference. Note, in the widely reported Fano resonances of photonic structures, it has been well recognized that the interference of a discrete state and a continuum state should give rise to a Fano lineshapes. This contradicts the Lorentzian lineshapes of the conventional waveguide-MRR. We ascribe it to the factor of phase-difference between the discrete state and the continuum state. In the conventional waveguide-MRR structure, the incident light splits into the MRR and the bus-waveguide at the coupling region, making them having the same initial phase. For the optical field forming resonant mode in the MRR, it experiences a phase delay of integer multiple of 2$\pi$ after it circulates back to the coupling region. Hence, when the discrete resonant modes couple back into the bus-waveguide, they have phase differences of integer multiple of 2$\pi$ relative to the continuum mode propagating directly in the bus-waveguide. According to the Fano-Anderson model, this 2$\pi$ phase-difference would result in a Fano lineshape with an infinite asymmetric factor $q$, which evolves into a symmetric Lorentzian lineshapes~\cite{miroshnichenko2005nonlinear}. This explains the Lorentzian resonance lineshapes presented in the transmission spectrum of a conventional waveguide-MRR structure. 

According to above analysis, to modify the Lorentzian resonance lineshapes of the MRR into asymmetric Fano lineshapes, it is essential to change the phase-difference between the discrete state and the continuum state from the integer multiple of 2$\pi$. To do that, we propose to add an air-hole in the bus-waveguide around the waveguide-MRR coupling region, as shown in Figs.~\ref{fig:fig1}(b). The inserted air-hole would induce a phase-shift for the continuum propagating mode in the bus-waveguide. On the other hand, since the air-hole does not modify the MRR structure, the condition of constructive interference in the MRR is unperturbed and the discrete resonant modes experience no extra phase-shift. Consequently, the phase difference between the discrete mode and the continuum mode is not equal to the integer multiple of 2$\pi$ any more, and asymmetric Fano resonance lineshapes are expected~\cite{miroshnichenko2005nonlinear}. The output optical field of the waveguide-MRR with an air-hole could be calculated by modifying Eq.~\ref{eq:1},

\begin{equation}
\begin{split}
E_{out}&=\underbrace{te^{-i\Delta\phi}{E_0}}_{\text{continuum state}}+\underbrace{i\kappa{E_1}+i\kappa{E_2}+\cdots}_{\text{discrete state}}\\
&=({te^{-i\Delta\phi}-\frac{\kappa^2ae^{i\delta}}{1-tae^{i\delta}})E_0}
\label{eq:3}
\end{split}
\end{equation}

Here, an extra phase-shift $\Delta\phi$ induced by the air-hole is added into the direct transmission part propagating in the bus-waveguide, which is considered as the continuum state. For the discrete state of the resonant modes, there is no difference from that in Eq.~\ref{eq:1}. The corresponding power transmission spectrum is calculated as 

\begin{equation}
T(\lambda)=\left|\frac{E_{out}}{E_{in}}\right|^2=\left|{te^{-i\Delta\phi}-\frac{\kappa^2ae^{i{2\pi{nL_R}/\lambda}}}{1-tae^{i{2\pi{nL_R}/\lambda}}}}\right|^2
\label{eq:4}
\end{equation}

\begin{figure}[http]
	\centering
	\fbox{\includegraphics[width=\linewidth]{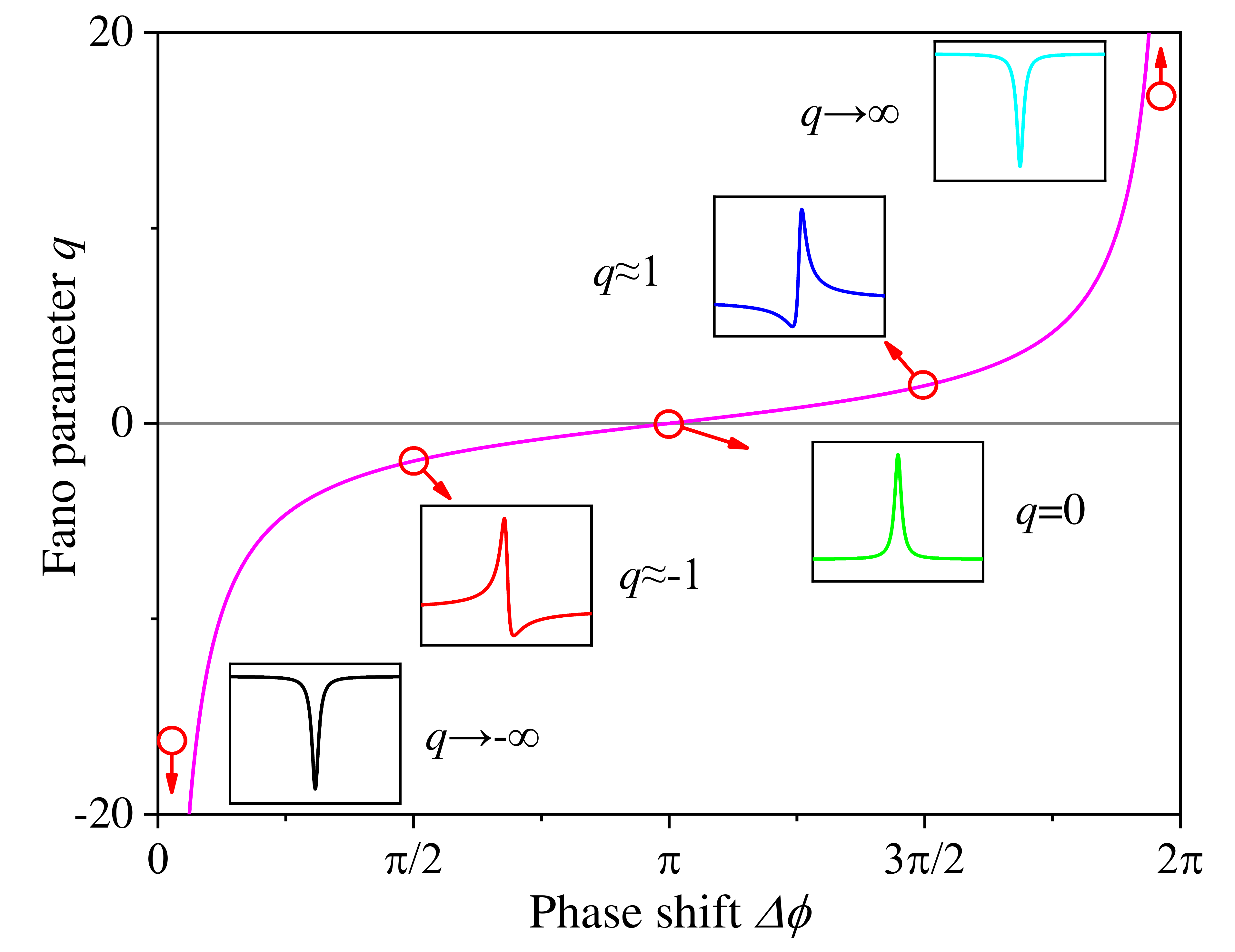}}
	\caption{Fano parameter $q$ is cotangent of the phase-shift $\Delta\phi$ with a period of 2$\pi$. The insets show the normalized Fano profiles at different $\Delta\phi$ as well as the extracted $q$ factors.}
	\label{fig:fig2}
\end{figure}

With the same parameters used to calculate Fig.~\ref{fig:fig1}(c) and by assuming $\Delta\phi=\pi/3$, the transmission spectrum of the waveguide-MRR with an air-hole is plotted in Fig.~\ref{fig:fig1}(d). Periodic Lorentzian lineshapes at the resonant wavelengths are modified into asymmetric Fano lineshapes successfully. To further illustrate the role of phase-shift $\Delta\phi$ played in the formation of Fano lineshapes, we calculate the transmission spectra of Eq.~\ref{eq:4} with different $\Delta\phi$, and fit them using Fano formula $T=1-(q+\Omega)^2/(1+\Omega^2)$~\cite{limonov2017fano, avrutsky2013linear}. Here, $\Omega$ is the reduced frequency and $q$ is the asymmetric parameter, which indicating the strengths of the discrete state and the continuum state during their coupling. In Fig.~\ref{fig:fig2}, the $q$ factors extracted from the Fano fittings are plotted with respect to the phase-shift $\Delta\phi$, presenting a cotangent-type function~\cite{yoon2013fano}. If the phase-shift equals to integer multiple of 2$\pi$, Eq.~\ref{eq:4} degenerates into Eq.~\ref{eq:2}, and the extracted $q$ factor tends to be infinity. In this case, because the constructive interference phenomena exists in the MRR, the discrete resonant modes couple with the continuum propagating mode  very weakly or negligibly. Therefore, the coupled Fano profile becomes symmetric Lorentzian function, as observed in the conventional waveguide-MRR. When $\Delta\phi=n\pi$ ($n$ is odd numbers), the Fano lineshape evolves into an electromagnetically induced transparency (EIT)-like peak, and $q=0$. Because of destructive interference between the discrete resonant mode and continuum propagating mode, it could be considered there is no coupling to the discrete state ($q=0$) and it describes the resonant suppression of the transmission~\cite{miroshnichenko2010fano}. If the phase-shift has other values, both constructive and destructive interference coexists between the discrete mode and the continuum mode, asymmetric Fano lineshapes are generated and the asymmetric parameters $q$ have non-zero values. Also, as shown in Fig.~\ref{fig:fig2}, the sign of $q$ alternates when $\Delta\phi$ changes with a step of $\pi$.

\section{Experiment Results}\label{sec:03}
\begin{figure}[http]
	\centering
	\fbox{\includegraphics[width=\linewidth]{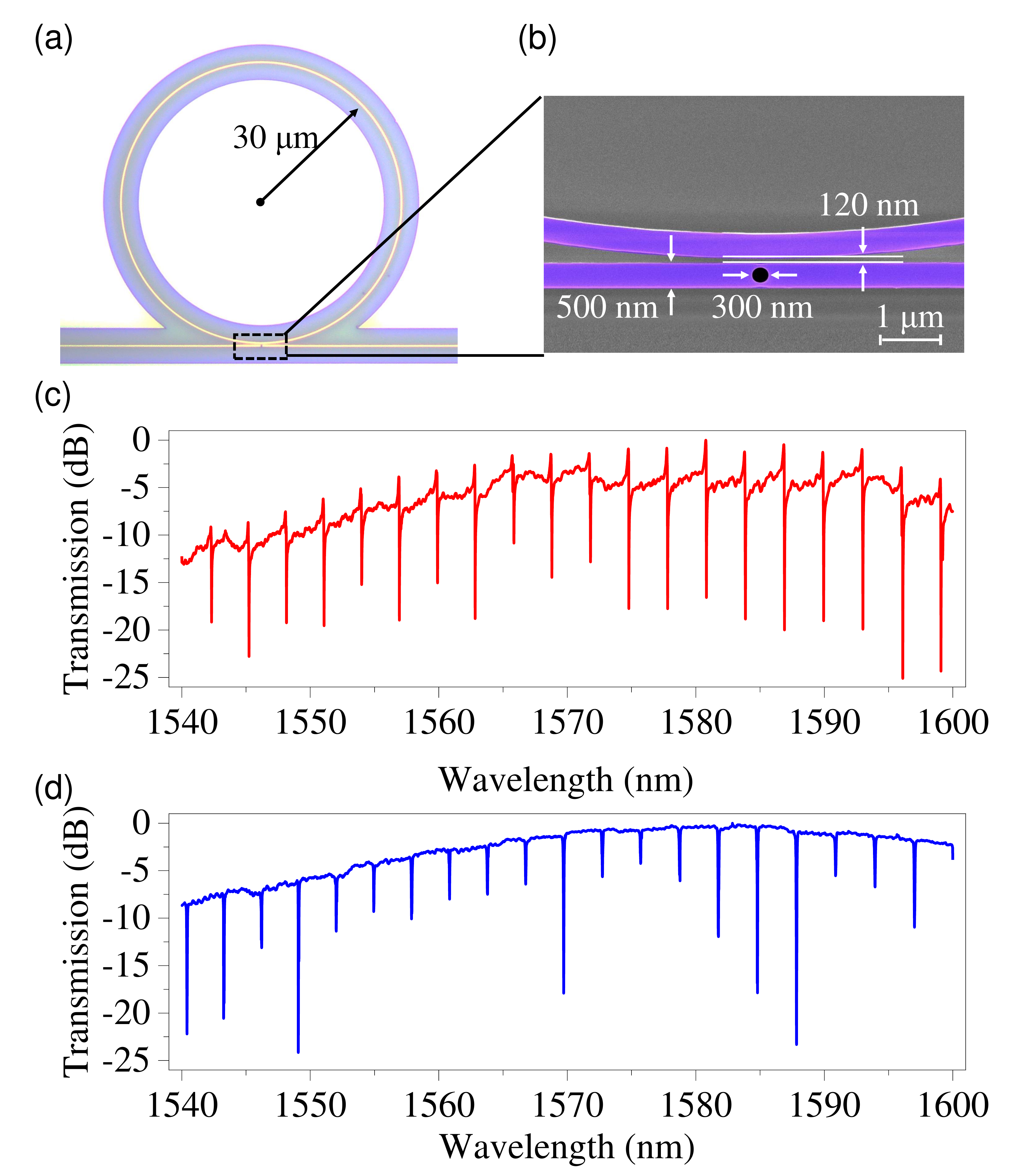}}
	\caption{(a) Optical microscope image and (b) the zoomed SEM image of the fabricated device. (c, d) (c) Transmission spectra of the device shown in (a). (d) Tranmission spectrum of conventional waveguide-MRR without an air-hole for the control experiment.}
	\label{fig:fig3}
\end{figure}

To verify the above theoretical predictions, we fabricate a side-coupled waveguide-MRR with an air-hole in the bus-waveguide, as displayed in the optical microscope image of Figure~\ref{fig:fig3}(a). The device is fabricated on a silicon-on-insulator chip, which has an 220 nm thick top silicon layer and a 2 $\upmu$m thick buried dioxide layer. Electron beam lithography and inductively coupled plasma etching are employed to define the devices. The coupling region of the waveguide-MRR is zoomed in a scanning electron microscope (SEM) image shown in Fig.~\ref{fig:fig3}(b). The waveguide and MRR have the same stripe width of 500 nm, and their coupling gap is 120 nm. The MRR radius is 30 $\upmu$m. The inserted air-hole has a diameter ($d$) of 300 nm and locates at the middle of the coupling region. 

We characterize this device by coupling a telecom-band narrowband tunable laser into the bus-waveguide via a vertical grating coupler. The transmitted optical power of the device is collected from an output vertical grating coupler, and measured by a photodiode. Figure~\ref{fig:fig3}(c) shows a measured transmission spectrum over a range from 1540 nm to 1600 nm. Fano resonance lineshapes are observed over a nearly flat transmission background with a spectral period determined by MRR's free spectral range (FSR~$\approx3$ nm). 
These Fano lineshapes show the largest extinction retios (ER) over 20 dB with a slope rate (SR) of 404.9 dB/nm (at 1599.03 nm). By fitting the lines to the Fano formula, the asymmetric factor and quality ($Q$) factor are extracted as $q\approx-1$ and $Q\approx31,000$, respectively. Analyzed from Eq.~\ref{eq:4}, the ER and SR could be improved by optimizing the waveguide-MRR structures~\cite{gu2019compact}. For example, similar as the ER in a conventional waveguide-MRR, by improving the fabrication process to match the critical coupling conditions, the ER could be  maximized. The SR could be improved by increasing the $Q$ factor, corresponding to narrow linewidth of the resonant mode. Also, while only the wavelength range of 60 nm is shown in the transmission spectrum, which is limited by the operation spectral range of the grating coupler, Fano lineshapes over an even broader spectral range should be observed by using butt-coupling geometry~\cite{ding2013ultrahigh}. We also implement a control experiment by fabricating a waveguide-MRR without air-hole in the bus-waveguide. The other parameters of the waveguide-MRR are not changed. The measured transmission spectrum is shown in Fig.~\ref{fig:fig3}(d). The resonant wavelengths and FSR are the same as those in the waveguide-MRR with air-hole. Differently, the resonant lineshapes are symmetric Lorentzian type. While the $Q$ factors of the resonant modes are evaluated exceeding 35,000, the SRs (around 100 dB/nm) are much lower than those obtained in the asymmetric Fano lineshapes. It could facilitate the Fano lineshapes for improving the performances of MRR-based switchings and sensors. 

\begin{figure}[http]
	\centering
	\fbox{\includegraphics[width=\linewidth]{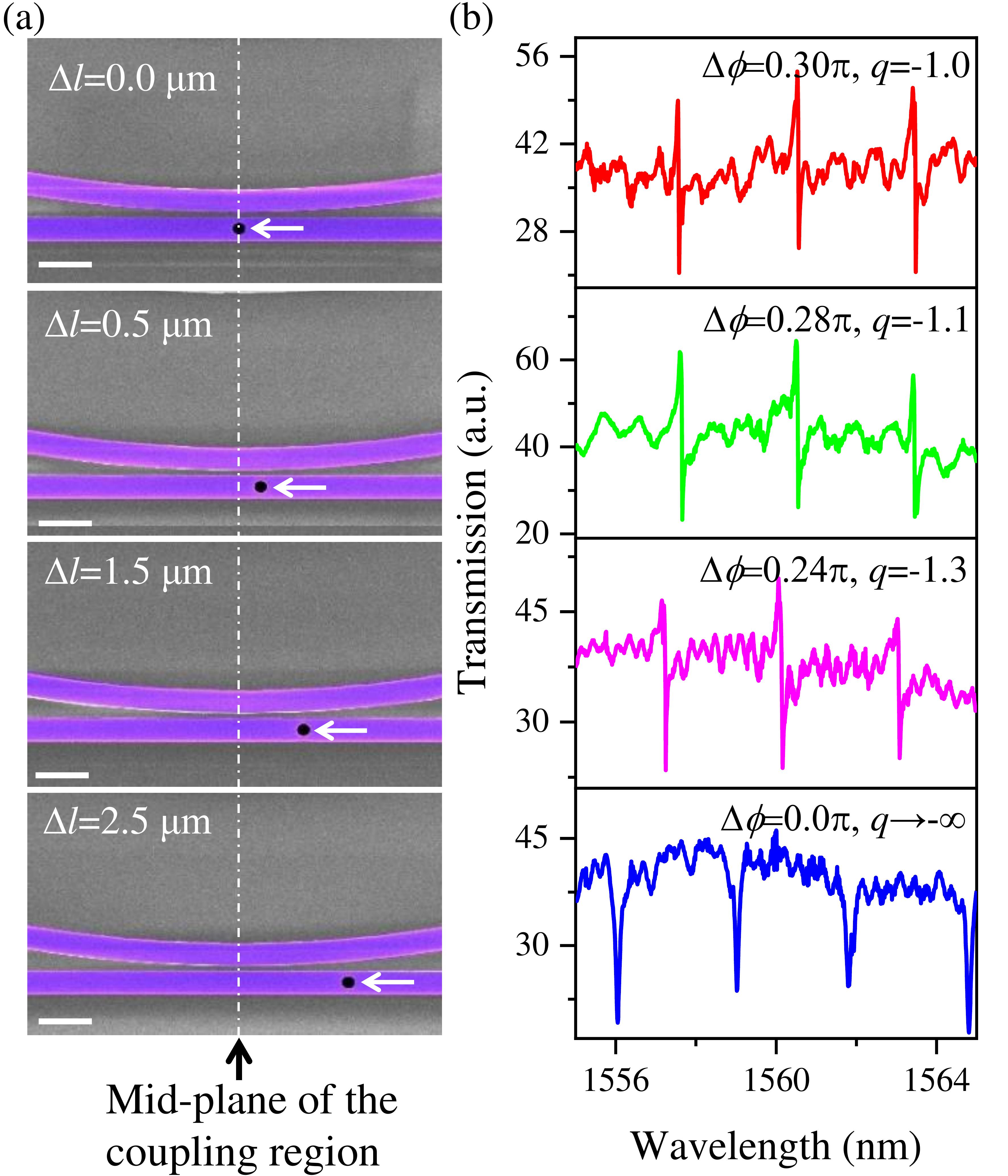}}
	\caption{ (a) SEM images and (b) transmission spectra of waveguide-MRR with the air-hole shifted from the mid-plane of the coupling region by $\Delta{l}$=0.0 $\upmu$m, 0.5 $\upmu$m, 1.5 $\upmu$m and 2.5 $\upmu$m. The scale bar corresponds to 1 $\upmu$m.}
	\label{fig:fig4}
\end{figure}

As analysed for Eq.~\ref{eq:3}, the asymmetric Fano lineshapes result from the air-hole-induced phase-difference between MRR's discrete resonant mode and bus-waveguide's continuum propagating mode. To further discuss this point, we experimentally fabricate more waveguide-MRR devices by varying the parameters of the inserted air-hole. First, based on the device design shown in Fig.~\ref{fig:fig3}(a), we move the air-hole away from the mid-plane of the waveguide-MRR coupling region gradually. SEM images of the fabricated devices are displayed in Fig.~\ref{fig:fig4}(a), which have air-hole location offsets ($\Delta{l}$) apart from the mid-plane of 0.0 $\upmu$m, 0.5 $\upmu$m, 1.5 $\upmu$m, 2.5 $\upmu$m.
\begin{figure}[http]
	\centering
	\fbox{\includegraphics[width=\linewidth]{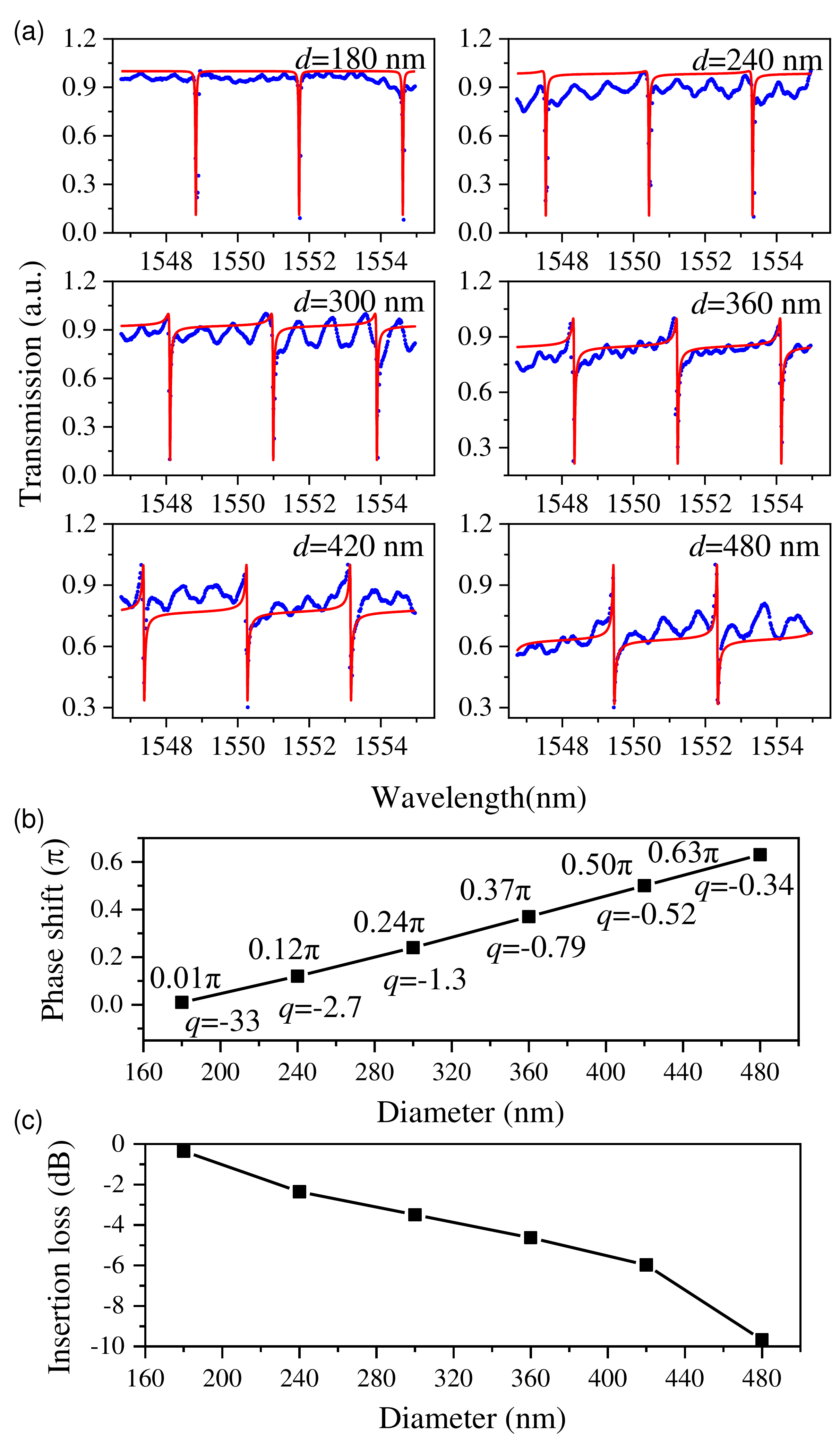}}
	\caption{(a) Measured (blue dots) and fitted (red solid, by Eq.~\ref{eq:4}) transmission spectra of devices with varied air-hole diameters. (b) Phase-shift and $q$ derived from the fitting in (a). (c) The measured insertion losses caused by different air-holes.}
	\label{fig:fig5}
\end{figure}
Transmission spectra of the devices are shown in Fig.~\ref{fig:fig4}(b) correspondingly. While the Fano lineshapes in each device are similar for all of the resonant modes, the resonant lineshapes vary obviously among different devices due to the varied $\Delta{l}$. By fitting these resonant lineshapes with Eq.~\ref{eq:4}, the phase shift $\Delta\phi$ and asymmetric factors $q$ are evaluated for each device, as shown in Fig.~\ref{fig:fig4}(b). The dependences of lineshapes on the air-hole locations prove the essential of phase-shift induced by the air-hole. For instance, for the device with $\Delta{l}=2.5$ $\upmu$m, the air-hole moves out of the waveguide-MRR coupling region, i.e., there is no air-hole in the coupling region. Hence, the discrete resonant modes in MRR coupling back to the bus-waveguide would interference with the unperturbed continuum propagating mode in the bus-waveguide, which maintains their phase-difference of integer multiple of 2$\pi$. After the coupling region, while there is an air-hole in the bus-waveguide, the interfering modes propagate together and would experience the same phase-shift. The transmitted resonant lineshapes have no difference from that obtained in the conventional waveguide-MRR, showing symmetric Lorentzian lineshapes. 

Next, devices with different air-hole diameters are fabricated and characterized, as shown in Fig.~\ref{fig:fig5}(a). The measured transmission spectra are fitted by Eq.~\ref{eq:4} and the $q$ factor and phase-shift $\Delta\phi$ induced by the air-holes are evaluated as well, as shown in Fig.~\ref{fig:fig5}(b). As the sizes of air-hole increase, phase-shifts and the asymmetric factor $q$ become larger. This is consistent with that effective refractive index of the propagating mode in the bus-waveguide is changed greater by the larger air-hole, resulting in larger phase-shift as well. Note, while the inserted air-hole induces a phase-shift of the bus-waveguide successfully, it is also a scattering point for the propagating mode. To facilitate potential applications of the air-hole-enabled Fano lineshapes, it is valuable to evaluate the inserting loss induced by the air-hole scattering. By comparing to the waveguide-MRR without air-holes, the insertion losses caused by the air-holes with different diameters are measured, as plotted in Fig.~\ref{fig:fig5}(c). For the air-hole with a diameter of 360 nm, which provides a remarkable Fano lineshape, the resulted insertion loss is about 5 dB. 

As predicted in Fig.~\ref{fig:fig2}, more plentiful lineshapes could be obtained if the phase-shifts are large enough, such as an EIT-like lineshape ($\Delta\phi=\pi$), Fano lineshapes with different asymmetric directions ($\Delta\phi>\pi$). However, from the above fabricated devices, the obtained maximal phase shift is about 0.63 $\pi$ with the air-hole diameter of 480 nm. This air-hole size approaches to the full width of the waveguide. Hence, it is impossible to achieve a phase shift of $\pi$ by inserting the circular air-hole. Also, as shown in Fig.~\ref{fig:fig5}(c), this air-hole with a diameter of 480 nm induces an insertion loss larger than 10 dB. To realize more lineshapes, the structure of the inserted air-hole could be optimized. For example, an elliptical air-hole or an array of small circular air-holes could be involved around the waveguide-MRR coupling region, which would induce larger phase shift in the bus-waveguide and reduce the insertion loss simultaneously. 

\section{Conclusion}\label{sec:04}
In conclusion, we have demonstrated that Fano resonance lineshapes could be realized reliably in a waveguide-MRR structure by simply inserting an air-hole in the side-coupled waveguide. By analysing the light propagation in the coupled structure, the air-hole-induced phase-shift is revealed to account for the formation of Fano resonance lineshapes. The theoretical predictions are verified experimentally well by fabricating devices with different parameters of the inserted air-holes. Fano resonance lineshapes are obtained over a broad spectral range with high ERs (> 20 dB) and SRs (> 400 dB/nm). Because the proposed design only requires to modify the bus-waveguide with an air-hole, eliminating the complex integration of other photonic structures, it reserves the compactness of the conventional waveguide-MRR structure. In addition, as discussed in the devices with different parameters of the air-hole, the generated Fano lineshapes have large tolerances of the structure designs and fabrications. We expect the air-hole-enabled Fano resonance lineshapes in the waveguide-MRR could be employed to improve the performances of MRR-based optical switchings, sensors and filters. 

~\\
\section*{Fundings}\label{sec:05}
National Natural Science Foundations of China (61775183, 11634010, 61522507); the Key Research and Development Program (2017YFA0303800, 2018YFA0307200); the Key Research and Development Program in Shaanxi Province of China (2017KJXX-12, 2018JM1058); the Fundamental Research Funds for the Central Universities (3102018jcc034, 3102017jc01001).

~\\
\section*{Acknowledgment}\label{sec:06}
The authors would thank the Analytical \& Testing Center of NPU for the assistances of device fabrication.

\bibliographystyle{plain}

\begin{thebibliography}{1}
	\bibitem{del2007optical} P. Del'Haye, A. Schliesser, O. Arcizet, T. Wilken, R. Holzwarth, and T. J. Kippenberg, “Optical frequency comb generation from a monolithic microresonator,” Nature 450, 1214 (2007).
	\bibitem{guo2017parametric} X. Guo, C.-l. Zou, C. Schuck, H. Jung, R. Cheng, and H. X. Tang, “Parametric down-conversion photon-pair source on a nanophotonic chip,” Light. Sci. \& Appl. 6, e16249 (2017).
	\bibitem{xu2005micrometre} Q. Xu, B. Schmidt, S. Pradhan, and M. Lipson, “Micrometre-scale silicon electro-optic modulator,” nature. 435, 325 (2005).
	\bibitem{kippenberg2011microresonator} T. J. Kippenberg, R. Holzwarth, and S. A. Diddams, “Microresonatorbased optical frequency combs,” science. 332, 555–559 (2011).
	\bibitem{wang2016fano} G. Wang, A. Shen, C. Zhao, L. Yang, T. Dai, Y. Wang, Y. Li, X. Jiang, and J. Yang, “Fano-resonance-based ultra-high-resolution ratio-metric wavelength monitor on silicon,” Opt. letters 41, 544–547 (2016).
	\bibitem{zhao2016tunable} G. Zhao, T. Zhao, H. Xiao, Z. Liu, G. Liu, J. Yang, Z. Ren, J. Bai, and Y. Tian, “Tunable fano resonances based on microring resonator with feedback coupled waveguide,” Opt. express 24, 20187–20195 (2016).
	\bibitem{qiu2012asymmetric} C. Qiu, P. Yu, T. Hu, F. Wang, X. Jiang, and J. Yang, “Asymmetric fano resonance in eye-like microring system,” Appl. Phys. Lett. 101, 021110 (2012).
	\bibitem{hu2013tunable} T. Hu, P. Yu, C. Qiu, H. Qiu, F. Wang, M. Yang, X. Jiang, H. Yu, and J. Yang, “Tunable fano resonances based on two-beam interference in microring resonator,” Appl. physics letters 102, 011112 (2013).
	\bibitem{mario2006asymmetric} L. Y. Mario, S. Darmawan, and M. K. Chin, “Asymmetric fano resonance and bistability for high extinction ratio, large modulation depth, and low power switching,” Opt. express 14, 12770–12781 (2006).
	\bibitem{tu2017high} Z. Tu, D. Gao, M. Zhang, and D. Zhang, “High-sensitivity complex refractive index sensing based on fano resonance in the subwavelength grating waveguide micro-ring resonator,” Opt. express 25, 20911–20922 (2017).
	\bibitem{chao2003biochemical} C.-Y. Chao and L. J. Guo, “Biochemical sensors based on polymer microrings with sharp asymmetrical resonance,” Appl. Phys. Lett. 83, 1527–1529 (2003).
	\bibitem{miroshnichenko2010fano} A. E. Miroshnichenko, S. Flach, and Y. S. Kivshar, “Fano resonances in nanoscale structures,” Rev. Mod. Phys. 82, 2257 (2010).
	\bibitem{absil2000compact} P. Absil, J. Hryniewicz, B. Little, R. Wilson, L. Joneckis, and P.-T. Ho, “Compact microring notch filters,” IEEE Photonics Technol. Lett. 12, 398–400 (2000).
	\bibitem{zhou2007fano} L. Zhou and A. W. Poon, “Fano resonance-based electrically reconfigurable add-drop filters in silicon microring resonator-coupled machzehnder interferometers,” Opt. letters 32, 781–783 (2007).
	\bibitem{darmawan2005phase} S. Darmawan, Y. Landobasa, and M. Chin, “Phase engineering for ring enhanced mach-zehnder interferometers,” Opt. express 13, 4580–4588 (2005).
	\bibitem{zhang2016optically} W. Zhang, W. Li, and J. Yao, “Optically tunable fano resonance in a grating-based fabry–perot cavity-coupled microring resonator on a silicon chip,” Opt. letters 41, 2474–2477 (2016).
	\bibitem{liang2006transmission} W. Liang, L. Yang, J. K. S. Poon, Y. Huang, K. J. Vahala, and A. Yariv, “Transmission characteristics of a fabry-perot etalon-microtoroid resonator coupled system,” Opt. letters 31, 510–512 (2006).
	\bibitem{miroshnichenko2005nonlinear} A. E. Miroshnichenko, S. F. Mingaleev, S. Flach, and Y. S. Kivshar, “Nonlinear fano resonance and bistable wave transmission,” Phys. Rev. E 71, 036626 (2005).
	\bibitem{limonov2017fano} M. F. Limonov, M. V. Rybin, A. N. Poddubny, and Y. S. Kivshar, “Fano resonances in photonics,” Nat. Photonics 11, 543 (2017).
	\bibitem{avrutsky2013linear} I. Avrutsky, R. Gibson, J. Sears, G. Khitrova, H. Gibbs, and J. Hendrickson, “Linear systems approach to describing and classifying fano resonances,” Phys. Rev. B 87, 125118 (2013).
	\bibitem{yoon2013fano} J. W. Yoon and R. Magnusson, “Fano resonance formula for lossy two-port systems,” Opt. express 21, 17751–17759 (2013).
	\bibitem{gu2019compact} L. Gu, H. Fang, J. Li, L. Fang, S. J. Chua, J. Zhao, and X. Gan, “A compact structure for realizing lorentzian, fano, and electromagnetically induced transparency resonance lineshapes in a microring resonator,” Nanophotonics 8, 841–848 (2019).
	\bibitem{ding2013ultrahigh} Y. Ding, H. Ou, and C. Peucheret, “Ultrahigh-efficiency apodized grating coupler using fully etched photonic crystals,” Opt. letters 38, 2732–2734 (2013).
\end{thebibliography}

\end{document}